\documentclass[twocolumn]{aastex62}

\usepackage{txfonts}
\usepackage{longtable}
\usepackage{rotating}
\usepackage{natbib}
\usepackage{graphicx}
\usepackage{graphics}
\usepackage{psfrag}
\usepackage{amssymb}
\bibpunct{(}{)}{;}{a}{}{,}
\def\Teff{$T_{\mathrm{eff}}$}

\submitjournal{ApJL}

\shorttitle{XUV Radiation from A-stars}
\shortauthors{Fossati et al.}

\begin{document}

\title{XUV Radiation from A-stars: Implications for Ultra-hot Jupiters}

\correspondingauthor{Luca Fossati}
\email{luca.fossati@oeaw.ac.at}

\author[0000-0003-4426-9530]{L. Fossati}
\affiliation{Space Research Institute, Austrian Academy of Sciences, Schmiedlstrasse 6, A-8042 Graz, Austria}

\author{T. Koskinen}
\affiliation{Lunar and Planetary Laboratory, University of Arizona, 1629 East University Boulevard, Tucson, AZ 85721-0092, USA}

\author{J. D. Lothringer}
\affiliation{Lunar and Planetary Laboratory, University of Arizona, 1629 East University Boulevard, Tucson, AZ 85721-0092, USA}

\author{K. France}
\affiliation{Laboratory for Atmospheric and Space Physics, University of Colorado, 600 UCB, Boulder, CO 80309, USA}
\affiliation{Center for Astrophysics and Space Astronomy, University of Colorado, 389 UCB, Boulder, CO 80309, USA.}

\author{M. E. Young}
\affiliation{Space Research Institute, Austrian Academy of Sciences, Schmiedlstrasse 6, A-8042 Graz, Austria}

\author{A. G. Sreejith}
\affiliation{Space Research Institute, Austrian Academy of Sciences, Schmiedlstrasse 		6, A-8042 Graz, Austria}








\begin{abstract}

\noindent Extremely irradiated, close-in planets to early-type stars might be prone to strong atmospheric escape. We review the literature showing that X-ray-to-optical measurements indicate that for intermediate-mass stars (IMS) cooler than $\approx$8250\,K, the X-ray and EUV (XUV) fluxes are on average significantly higher than those of solar-like stars, while for hotter IMS, because of the lack of surface convection, it is the opposite. We construct spectral energy distributions for prototypical IMS, comparing them to solar. The XUV fluxes relevant for upper planet atmospheric heating are highest for the cooler IMS and lowest for the hotter IMS, while the UV fluxes increase with increasing stellar temperature. We quantify the influence of this characteristic of the stellar fluxes on the mass loss of close-in planets by simulating the atmospheres of planets orbiting EUV-bright (WASP-33) and EUV-faint (KELT-9) A-type stars. For KELT-9b, we find that atmospheric expansion caused by heating due to absorption of the stellar UV and optical light drives mass-loss rates of $\approx$10$^{11}$\,g\,s$^{-1}$, while heating caused by absorption of the stellar XUV radiation leads to mass-loss rates of $\approx$10$^{10}$\,g\,s$^{-1}$, thus underestimating mass loss. For WASP-33b, the high XUV stellar fluxes lead to mass-loss rates of $\approx$10$^{11}$\,g\,s$^{-1}$. Even higher mass-loss rates are possible for less massive planets orbiting EUV-bright IMS. We argue that it is the weak XUV stellar emission, combined with a relatively high planetary mass, which limit planetary mass-loss rates, to allow the prolonged existence of KELT-9-like systems.

\end{abstract}

\keywords{planets and satellites: atmospheres --- planets and satellites: gaseous planets --- planets and satellites: individual (KELT-9b, WASP-33b) --- stars: early-type --- stars: activity}

%
\section{Introduction} \label{sec:introduction}
\noindent
Hot Jupiters (HJs) are subject to mass loss by hydrodynamic escape \citep[e.g.,][]{vidal2003,fossati2010}. Hydrogen and helium in the upper atmosphere absorb high-energy and ionizing (X-ray and EUV, together called XUV) stellar radiation, leading to atmospheric heating and expansion, thus mass loss \citep{yelle2004}. Both approximations \citep{erkaev2007,kubyshkina2018} and detailed hydrodynamic modelling \citep[e.g.,][]{koskinen2013a,koskinen2013b,koskinen2014} concur on values of the order of 10$^{9-10}$\,g\,s$^{-1}$ for the mass-loss rates of classical HJs, e.g., HD209458b and HD189733b.

Atmospheric escape is a fundamental process affecting planetary atmospheric structure, composition, and evolution. For HJs, \citet{szabo2011} and \citet{mazeh2016} presented evidence for a lack of planets with orbital periods shorter than 1-2\,days and masses smaller than one Jupiter mass ($M_{\rm J}$) in the population of known exoplanets (i.e., sub-Jovian desert). \citet{owen2018} suggested that this might be the result of the interplay between atmospheric escape and planet migration.

Most exoplanets known to date orbit late-type stars, which are characterized by the presence of a chromosphere ($\approx$10$^4$\,K), a transition region ($\approx$10$^5$\,K), and a corona ($\approx$10$^6$\,K) lying on top of the photosphere. The XUV flux of late-type stars originates in these layers. There is, however, a rapidly increasing number of extremely irradiated HJs being detected around intermediate-mass stars (IMS; main-sequence F5- to B5-type stars). Examples of such systems are WASP-33b, KELT-20b, WASP-189b, MASCARA-1b, and KELT-9b \citep{collier2010,lund2017,anderson2018,talens2017,gaudi2017}. The \textit{TESS} mission is also expected to find several more of these systems \citep{barclay2018}. Adequately estimating the XUV flux emitted by IMS is therefore critical for the determination of the physical characteristics and evolution of these extremely irradiated planets.

We review observational evidence indicating that the XUV flux of hotter IMS provide a significantly different environment to their planets (Sect.~\ref{sec:xuv}) and use the information available to estimate the XUV fluxes of different IMS. We then discuss the impact of these XUV fluxes on the atmosphere of HJs orbiting IMS (Sect.~\ref{sec:implications}) and present our conclusions in Sect.~\ref{sec:conclusions}.
\section{The high-energy fluxes of intermediate-mass stars}\label{sec:xuv}
\noindent While there is a vast literature on the XUV fluxes of late-type stars, much less is known about those of IMS. Stellar models suggest that the surface convection zone thins with increasing effective temperature (\Teff) up to $\approx$8500\,K (i.e., $B-V$\,$\approx$\,0.1; spectral type A3), at which point it vanishes rapidly \citep[e.g.,][]{bohm1984,christensen2000,kupka2002,samadi2002}. This theoretical prediction finds support in multi-wavelength observations.

\begin{figure}[h!]
\includegraphics[width=\hsize,clip]{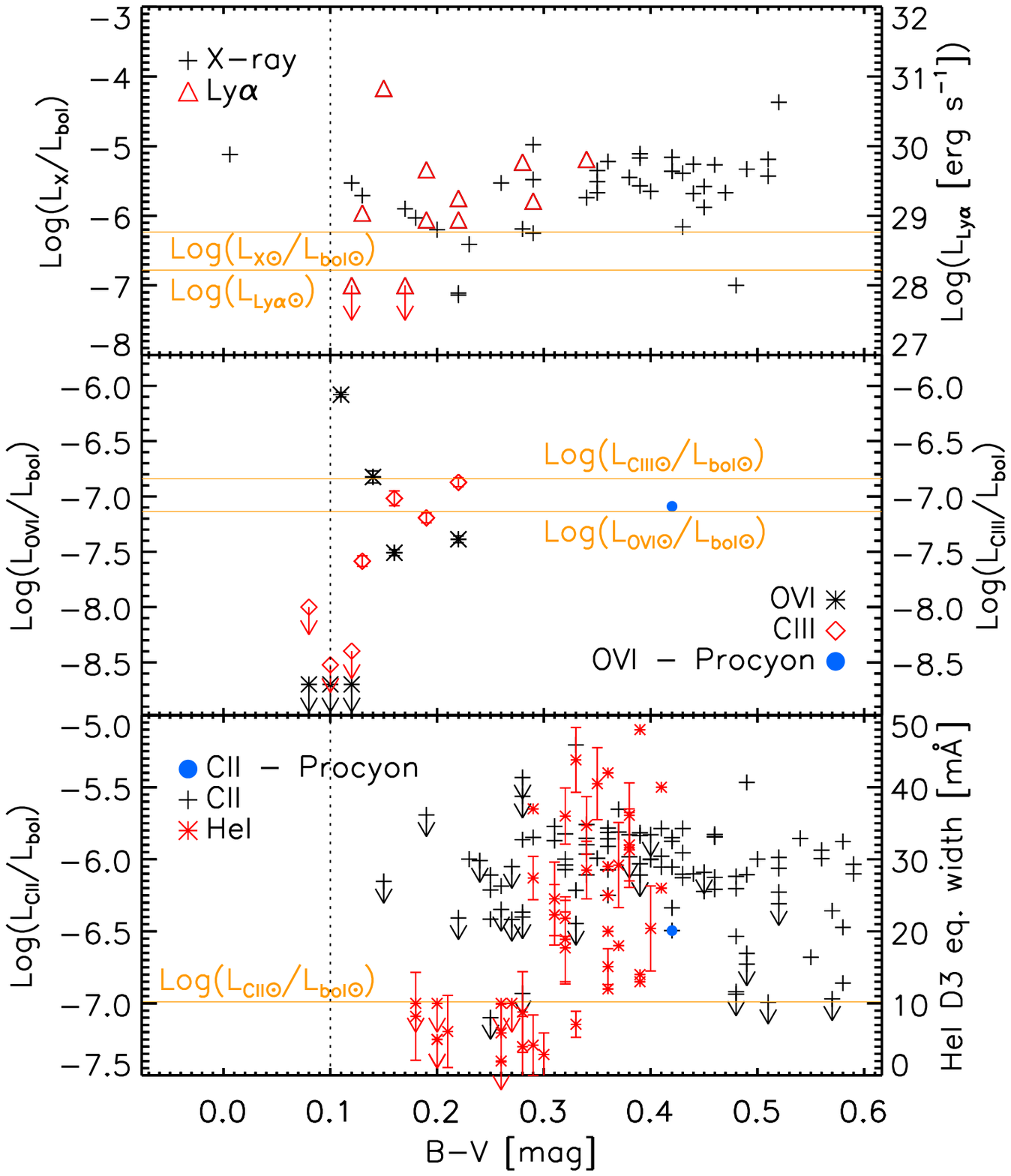}
\caption{Top: X-ray luminosity \citep[black plus sign;][]{panzera1999} and Ly$\alpha$ luminosity \citep[red triangle;]{marilli1997} vs stellar $B-V$ color. The Ly$\alpha$ luminosities are not corrected for the bolometric luminosity because of lack of data. Several early A-type stars are detected in the X-rays, but the emission comes from low-mass companions. Middle: O{\sc vi} \citep[black asterisk;][]{neff2008} and C{\sc iii} \citep[red rhombi;][]{simon2002} luminosity vs $B-V$ color. Bottom: C{\sc ii} luminosity \citep[black plus sign;][]{simon1991} and equivalent width of the He{\sc i}\,D$_3$ line \citep[red asterisk;][]{rachford1997} vs $B-V$ color. Downward arrows indicate upper limits. Horizontal lines mark solar values. The vertical line indicates the $B-V$ color at which stellar activity is believed to fade. In the middle and bottom panels, the blue dots indicate Procyon's O{\sc vi} and C{\sc ii} luminosities. Ly$\alpha$, O{\sc vi}, C{\sc iii}, and C{\sc ii} emission and He{\sc i} absorption are undetected for stars earlier than A3.}
\label{fig:FUVmeasurements}
\end{figure}

\citet{simon1995}, \citet{panzera1999}, and \citet{schroder2007} analyzed {\it ROSAT} X-ray observations of a large sample of A-type stars. In general, they found an increasing number of X-ray detections with decreasing temperature. The results of \citet{panzera1999} are representative. They considered a sample of 66 A- and 76 F-type stars ranging between spectral types A0 and F6 detected with \textit{ROSAT}. By excluding binaries, for which the X-ray emission might come from the low-mass companion, they obtained a sample of 19 A- and 33 F-type stars showing X-ray emission, (almost) all later than spectral type A3 (Fig.~\ref{fig:FUVmeasurements}).

The \textit{EUVE} satellite collected EUV (70--760\,\AA) spectra of several nearby IMS, but detecting just Procyon (F5) and Altair \citep[A7; both also detected in X-rays; e.g.,][]{simon1995}, while no EUV emission was detected for Vega \citep[A0; also undetected in X-rays;][]{schroder2007}, which is the nearest hot IMS \citep{craig1997}. This indicates that Procyon and Altair have hot chromospheres and coronae, in contrast to Vega \citep{craig1997}.

Since EUV radiation is heavily absorbed by the interstellar medium, the search for and characterization of chromospheres and coronae on IMS can be better performed at far-ultraviolet (FUV; 912--1800\,\AA) wavelengths. This is because the FUV waveband contains emission lines with formation temperatures similar to those of the EUV emission (i.e., 10$^{5-6}$\,K). \citet{simon1991}, \citet{schrijver1993}, \citet{simon1994}, and \citet{walter1995} showed that the C{\sc ii} (1335\,\AA) emission flux of early F-type stars can be as high as that of the most active late-type dwarfs. Furthermore, Ly$\alpha$ emission is clearly detected in stars only up to spectral type A4 \citep[e.g.,][]{marilli1997,simon2002}.

\begin{figure}[h!]
\includegraphics[width=\hsize,clip]{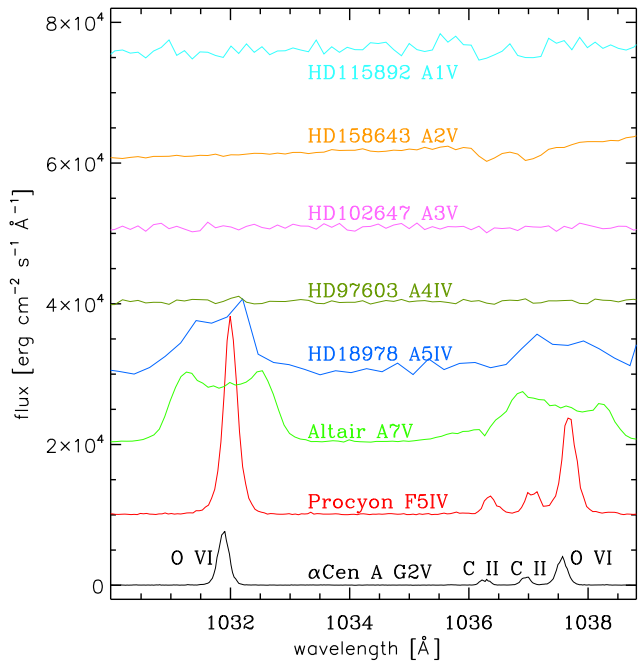}
\caption{\textit{FUSE} surface fluxes in the region of the C{\sc ii} 1036 and 1037\,\AA\ and of the O{\sc vi} 1032 and 1037\,\AA\ lines for G2- to A1-type stars. We applied rigid shifts and rebinning for visualization purposes. The O{\sc vi} emission disappears abruptly around spectral type A4--A5. The C{\sc ii} emission turns into absorption for the hotter stars. Altair and HD18978 are fast rotators, hence the broad emission lines.}
\label{fig:fuse}
\end{figure}

Observations at wavelengths longer than Ly$\alpha$ (1216\,\AA) are limited to the cooler IMS because the diagnostic lines (e.g., C{\sc ii} 1335\,\AA, Si{\sc iv} 1400\,\AA, C{\sc iv} 1550\,\AA) become harder to detect against the photospheric background rising with increasing \Teff. By lying at shorter wavelengths, the spectral region bluewards of Ly$\alpha$ (912--1200\,\AA) is less affected by stellar continuum emission up to the early A-type stars. \citet{simon2002} and \citet{neff2008} reported measurements of the chromospheric and transition region emission lines C{\sc iii} (977\,\AA), O{\sc vi} (1032 and 1037\,\AA), and C{\sc iii} (1175\,\AA) obtained from \textit{FUSE} spectra for a set of seven nearby IMS in the 7800--8600\,K \Teff\ range (i.e., A7--A2). They found strong emission lines for stars with \Teff\,$\leq$\,8200\,K (A4), beyond which no line emission is detected (Fig.~\ref{fig:fuse}). They also showed that the O{\sc vi} luminosity correlates tightly with the X-ray luminosity, as expected for an emission generated by hot chromospheric and coronal material.

It is also possible to gather information about chromospheric and coronal properties of IMS from observations in the optical band. The He{\sc i}\,D$_3$ (5876\,\AA) absorption line is a classical activity indicator, present in stars with chromospheres \citep[e.g.,][]{wolff1984}. \citet{garcia1993} and \citet{rachford1997} collected spectra for about 50 IMS concluding that the line appears to be present for stars later than A5 and that the line strength is independent of rotation and age.

Multi-wavelength measurements (Fig.~\ref{fig:FUVmeasurements}) concur on the presence of a sharp boundary at 8000--8500\,K, where cooler stars present chromospheres and coronae, thus high XUV fluxes, while hotter stars do not. These measurements also indicate that the XUV emission of cooler IMS is likely to be significantly stronger than that of inactive late-type stars like the Sun. This is confirmed by comparing the XUV fluxes of Procyon and solar, showing that the former are about three times higher than the latter \citep{fossati2018}.

\begin{figure}[h!]
\includegraphics[width=\hsize,clip]{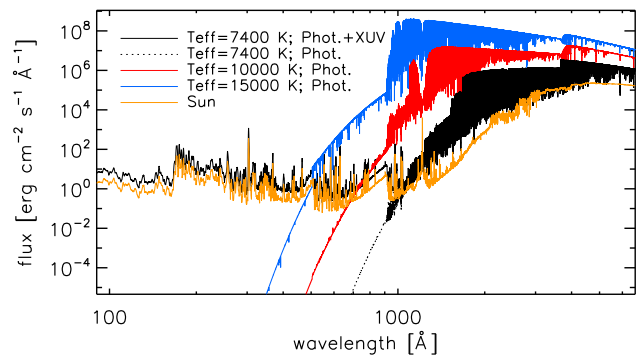}
\caption{Comparison between non-LTE SEDs computed for IMS with temperatures of 7400\,K (black) and 10000\,K (red), and for the Sun (orange) and a 15000\,K star (blue). For the cooler IMS, we used the XUV fluxes of Procyon at wavelengths below 1100\,\AA, while the photospheric fluxes are shown by the dashed line.  All fluxes are scaled to a distance of 0.03\,AU and account for the radii typical of stars with the respective temperatures.}
\label{fig:sed}
\end{figure}

To estimate the impact of the presence/lack of XUV irradiation on the atmosphere of close-in planets orbiting IMS, we computed spectral energy distributions (SED) for a 7400\,K star (similar to WASP-33) and a 10000\,K star (similar to KELT-9), thus for stars with and without a chromosphere/corona, employing the non-LTE \textit{PHOENIX} stellar model atmosphere code, v15 \citep[][Fig.~\ref{fig:sed}]{phoenix}. In this comparison, we further consider the Sun and a \Teff\,=\,15000\,K star because planets are in principle detectable with \textit{TESS} orbiting such high-temperature stars \citep{bouma2017}. We rescaled all fluxes to a distance of 0.03\,AU, i.e., in between the orbital separations of WASP-33b and KELT-9b \citep{collier2010,gaudi2017}. For the cooler IMS, we assumed that the XUV flux below 1100\,\AA\ is similar to that of Procyon. However, Fig.~\ref{fig:FUVmeasurements} indicates that Procyon's XUV flux might be underestimating that of a 7400\,K star. Table~\ref{tab:sed} quantifies the short-wavelength emission of the considered stars.

Based on the evidence presented here, cooler IMS have strong XUV fluxes at wavelengths shorter than the hydrogen ionization threshold (912\,\AA) that are relevant to heating the upper atmospheres of giant planets orbiting IMS, hereafter ultra-hot Jupiters (UHJs; Fig.~\ref{fig:sed}; Table~\ref{tab:sed}). In contrast, the XUV flux of hotter IMS decreases rapidly with decreasing wavelength, with $\approx$99\% of the hydrogen ionizing radiation confined between 750 and 912\,\AA, making the XUV fluxes below 750\,\AA, the more relevant for driving planet atmospheric escape (see below), negligible compared to those of the other stars. The hotter IMS, however, have the strongest FUV and near-UV (NUV; 1800--3200\,\AA) fluxes. We expect that this difference in the SEDs of IMS has interesting consequences for the heating and escape of UHJ.
\begin{table*}
\caption{Comparison between the integrated XUV, FUV, and NUV fluxes obtained for the stars shown in Fig.~\ref{fig:sed}. The integrated fluxes are in erg\,cm$^{-2}$\,s$^{-1}$ and are for a distance of 0.03\,AU. Values in parenthesis are for fluxes below 750\,\AA. The last two columns give flux ratios comparing the XUV and FUV or FUV$+$NUV.}
\label{tab:sed}
\begin{center}
\begin{tabular}{l|ccccc}
\hline
\hline
Star & XUV & FUV & NUV & $\frac{\rm XUV}{\rm FUV}$ & $\frac{\rm XUV}{\rm FUV+NUV}$ \\
     & $<$912\,\AA\  ($<$750\,\AA) & 912--1800\,\AA & 1800--3000\,\AA & & \\
     & [erg\,cm$^{-2}$\,s$^{-1}$] & [erg\,cm$^{-2}$\,s$^{-1}$] & [erg\,cm$^{-2}$\,s$^{-1}$] & & \\
\hline
10000\,K & 4000 (60)                            & 6$\times$10$^{9}$ & 1$\times$10$^{10}$ & 7$\times$10$^{-7}$ (1$\times$10$^{-8}$) & 2$\times$10$^{-7}$ (3$\times$10$^{-9}$) \\
7400\,K  & 7800 (6500)                          & 3$\times$10$^{7}$ & 1$\times$10$^{9}$  & 2$\times$10$^{-4}$ (2$\times$10$^{-4}$) & 7$\times$10$^{-6}$ (6$\times$10$^{-6}$) \\
\hline
Sun\,    & 2400 (2000)                          & 3$\times$10$^{4}$ & 3$\times$10$^{7}$  & 1$\times$10$^{-1}$ (8$\times$10$^{-2}$) & 8$\times$10$^{-5}$ (7$\times$10$^{-5}$) \\
15000\,K & 6$\times$10$^{6}$ (4$\times$10$^{5}$) & 2$\times$10$^{11}$ & 1$\times$10$^{11}$ & 3$\times$10$^{-5}$ (2$\times$10$^{-6}$) & 2$\times$10$^{-5}$ (1$\times$10$^{-6}$) \\
\hline
\end{tabular}
\end{center}
\end{table*}
%
\section{Implications for exoplanets}\label{sec:implications}
\noindent
The possible difference in XUV fluxes from cooler to hotter IMS would have profound implications for atmospheric modeling and interpretation of the observations of UHJs. This is especially true for upper atmosphere observations and mass-loss predictions. Current models of UHJ atmospheres show that stellar continuum fluxes produce high temperatures leading to the dissociation of molecules and thermal ionization of low-ionization threshold metals such as Si, Mg, Fe, Na, and K at relatively high pressures ($p$\,$>$\,1\,$\mu$bar) compared to classical HJs \citep{kitzmann2018,lothringer2018,parmentier2018}. In particular, only simple diatomic molecules (e.g., H$_2$, CO, OH) may survive in the upper atmosphere ($p$\,$<$\,1\,mbar) with detectable abundances.  

\citet{lothringer2018} and \citet{kitzmann2018} suggested that CO infrared bands would be detectable in the spectra of UHJs with \textit{JWST}. Given the high photospheric FUV fluxes of hot IMS, the FUV bands of CO \citep{CO} may also be detectable during primary transit. The alkali atom lines in the optical that are prominent features in the transmission spectrum of classical HJs, on the other hand, should be significantly muted by ionization around the 10--100\,mbar level. Metal opacities (Fe and Mg) and absorption from SiO, metal hydrates, and H$^-$, which is the main infrared opacity source in UHJs, lead to the formation of a temperature inversion in the lower atmosphere, which is present even without invoking TiO and VO absorption \citep{lothringer2018}.

KELT-9b is the hottest of the known UHJs and as such it presents a fascinating test case for mass-loss studies. The energy-limit for mass loss is given by \citep{erkaev2007}
\begin{equation}
\dot{M} = \frac{\pi \epsilon F_{\rm XUV} r^2}{\Phi_0 K} \left( \frac{R_{\rm s}}{a} \right)
\end{equation}
where $\epsilon$ is the heating efficiency, $F_{\rm XUV}$ is the stellar surface XUV flux, $\Phi_0$ is the gravitational potential at the planetary radius, $r$ is the effective radius in the atmosphere where the XUV flux is absorbed, $R_{\rm s}$ is the stellar radius, $a$ is the orbital semi-major axis, and $K$ accounts for Roche lobe effects and depends on the ratio of the L1 point radius to the planetary radius. Assuming the XUV flux value from Table~\ref{tab:sed} for a 10000\,K star, energy-limited escape from KELT-9b would be $\approx$10$^{11}$\,g\,s$^{-1}$. However, this value is based on 100\% heating efficiency, which is too high. To highlight this effect, we arbitrarily consider a reference wavelength of 750\,\AA\  (see Table~\ref{tab:sed}), which is the approximate position at which the fluxes of the star with \Teff\,=\,10000\,K cross those of the star with \Teff\,=\,7400\,K. Since the heating efficiency increases with photon energy up to a point where the photoelectron reaches the ionization potential of hydrogen, the heating efficiency for ionizing radiation at wavelengths longer than 750\,\AA\  (i.e., where most of the stellar XUV flux is emitted) is $\approx$10\%, thus the energy-limited mass-loss rate is probably closer to 10$^{10}$\,g\,s$^{-1}$. This result is independent of the considered reference wavelength, because at wavelengths longer than 750\,\AA\ the XUV fluxes are higher, but the heating efficiency is lower, while at shorter wavelengths it is the opposite. This is the result of the strong wavelength dependency of the XUV fluxes of hot IMS with wavelength, which is instead not the case for late-type stars. Given that detailed calculations addressing the contribution of the longer wavelength FUV radiation to the energy budget of the upper atmosphere do not exist, at this stage it is acceptable to assume an energy-limited escape rate of 10$^{10}$--10$^{11}$\,g\,s$^{-1}$.      

We note that the energy limit based on the XUV flux is not adequate for predicting mass-loss rates for UHJs due to the dissociation of molecules, including H$_2$, at deeper pressures than on classical HJs (Koskinen et al., in preparation). To illustrate this point, we use the model of \citet{lothringer2018} to estimate the mass-loss rate for KELT-9b. To do so, we calculate the extent of the atmosphere based on the predicted T-P profile and composition, including the effect of the Roche potential and assuming that energy is fully redistributed around the planet. We find that the pressure of the Roche lobe equipotential surface is about 10$^{-11}$\,bar. Given that the effective thermal escape parameter approaches zero at the Roche lobe boundary, Jeans escape reduces to the ``free particle'' (sound speed) limit and we find a mass-loss rate of 10$^{10}$--10$^{11}$\,g\,s$^{-1}$ i.e., comparable to the energy limit. This mass-loss rate is almost entirely powered by the expansion of the atmosphere due to the heating of the middle and upper atmosphere by UV and visible radiation, while XUV thermospheric heating has only a small effect on the T-P profile.      
 
\citet{casasayas2018} presented results of high-resolution optical transmission spectroscopy of the UHJ MASCARA-2b, including the detection of Na and H$\alpha$ absorption. Hydrogen absorption at the position of the H$\alpha$ line has been detected in transmission also for the UHJ KELT-9b \citep{yan2018}. Both of these planets orbit IMS with temperatures well above the 8000--8500\,K threshold. We used the model results from \citet{lothringer2018} to estimate the magnitude of KELT-9b H$\alpha$ absorption, assuming local thermodynamical equilibrium (LTE) level populations for hydrogen, similar to \citet{yan2018} and \citet{casasayas2018}. We find that the abundance of the $n$\,=\,2 population peaks around 0.2\,$\mu$bar, with a mixing ratio of about 10$^{-8}$, in line with the relatively high temperature of about 6000\,K at that pressure level. This leads to a H$\alpha$ core transit depth of about 1.2\%.

Our results show that a significant excess transit depth over the continuum transit depth of $\approx$0.7\% can be produced without invoking strong additional mass loss by XUV radiation. This conclusion agrees qualitatively with the results of \citet{casasayas2018} for MASCARA-2b. Our calculated transit depth of KELT-9b, however, falls short of the observed H$\alpha$ transit depth of 1.8\% \citep{yan2018}. There are several possible reasons for this discrepancy. The atmosphere could be hotter and more extended than we predicted, leading to a larger transit depth and higher mass-loss rate. The significant mass uncertainty \citep[2.88$\pm$0.84\,$M_{\rm J}$][]{gaudi2017} can also contribute to the discrepancy. If the true planetary mass is closer to the lower-end of the 1$\sigma$ range, the atmosphere would be more extended with a larger transit depth and higher mass-loss rate, without the need to increase the temperature.  

It would be possible to explore different temperature profiles and planet masses to obtain a better fit with the observations, but this exercise, however, is not particularly informative at this point, for two main reasons. First, the current uncertainties in the system properties preclude strong conclusions on mass loss based on the existing observations. Second, the hydrogen $n$\,=\,2 level population in the upper atmosphere likely deviates significantly from LTE.  Detailed calculations of level populations have been undertaken on classical HJs \citep{menager2013,huang2017}, but have not been extended to UHJs and are out of scope for this letter. They do indicate, however, that the inclusion of radiative and photoelectron excitation could potentially enhance the $n$\,=\,2 level population.

\citet{yan2018} employed the information derived from the depth of the H$\alpha$ line of KELT-9b to infer a planetary mass-loss rate of about 10$^{12}$\,g\,s$^{-1}$, assuming Jeans escape. They then mentioned that this value might be underestimated, because it ignores escape driven by the stellar XUV flux. In fact, both \citet{casasayas2018} and \citet{yan2018}, who regarded hot IMS as if they would have XUV fluxes stronger than those of solar-like stars, interpreted the H$\alpha$ absorption as being the result of excitation due to XUV absorption. Instead, we argue that the mass-loss rate derived by \citet{yan2018} could be close to correct or even higher than the real mass-loss rate, because the star lacks strong emission at the wavelengths most relevant for driving escape that is not necessarily required to explain the existing transit observations.    

As indicated by Table~\ref{tab:sed}, UHJs orbiting cool IMS, such as WASP-33b, are instead probably subject to XUV irradiation significantly larger than that of classical HJs. The EUV flux is absorbed by hydrogen and helium in the upper atmosphere while X-rays and UV radiation are absorbed by metals in the lower atmosphere. This implies that planets orbiting cool IMS could undergo significant heating across the whole atmosphere that will lead to powerful hydrodynamic escape, possibly one or more orders of magnitude larger than that of classical HJs. This, however, does not appear to be the case for WASP-33b, where the energy-limited mass-loss rate is $\approx$10$^{11}$\,g\,s$^{-1}$ based on the XUV flux given in Table~\ref{tab:sed} and heating in the middle atmosphere does not add significantly to the mass-loss rate. In contrast to planets orbiting cooler IMS, planets orbiting hot IMS may be subject almost exclusively to UV irradiation, which is absorbed by metals. As a result, the available energy to heat the thermosphere is much smaller and mass-loss rates are suppressed unless the planet is also subject to significant Roche lobe overflow. From the planet's perspective, the most threatening scenario is Roche lobe overflow partly powered by UV and visible radiation combined with significant XUV fluxes, which we argue is not the case for KELT-9b, providing a compelling explanation for the detection of a planet in such an extreme environment in the first place.        
\section{Conclusions}\label{sec:conclusions}
\noindent We reviewed past X-ray, EUV, UV, and optical observations of IMS showing the split nature of their XUV emission. For the cooler IMS, the XUV fluxes, particularly those most relevant for driving planet atmospheric escape ($<$750\,\AA), are on average significantly higher than those of solar-like stars, while for the hotter IMS later than mid B-type it is the opposite. Observations concur at setting the threshold at $\approx$8250\,K.

We estimated the impact of the XUV fluxes peculiar behavior on the atmospheric escape of UHJs. We concluded that the energy-limited approach alone is not applicable to many UHJs, as it may underestimate escape rates. For KELT-9b, which is subject to low XUV, but high UV irradiation levels, we found XUV-driven mass-loss rates of $\approx$10$^{10}$--10$^{11}$\,g\,s$^{-1}$, while atmospheric expansion caused by heating due to absorption of the stellar UV and optical light alone leads to similar mass-loss rates. For UHJs orbiting cooler IMS, such as WASP-33b, the strong XUV stellar fluxes power mass-loss rates of $\approx$10$^{11}$\,g\,s$^{-1}$, though they could be even higher because of the additional heating of the middle atmosphere.

The weak emission at wavelengths shorter than 750\,\AA\ for the hotter IMS and the relatively high mass of the planet may explain why the atmosphere of KELT-9b hasn't completely escaped. Therefore, the planet should not be in a short-lived phase of evolution, hence more KELT-9b-like planets might exist. The generally high mass-loss rates we estimated for UHJs indicate that atmospheric escape plays a pivotal role in shaping their evolution and that UHJs are ideal laboratories to study escape.

The stellar EUV fluxes are a key ingredient for future mass-loss studies of UHJs. However, observations at EUV wavelengths have not been possible since the decommissioning of the \textit{EUVE} satellite. Additionally, the modest aperture and low-throughput of \textit{EUVE} limited it to only the nearest low- and intermediate-mass stars. A new EUV mission featuring higher sensitivity (effective area $>$50\,cm$^2$)\footnote{The peak \textit{EUVE} effective area was 2\,cm$^2$ ({\tt https://archive.stsci.edu/euve/handbook/handbook.html}).} could overcome these prior limitations to survey EUV emission from a statistical sample of stars within 30\,pc, placing exoplanet mass loss observations on a firmer empirical basis.

%
\acknowledgments
\noindent LF and AGS acknowledge support from FFG-P859718. We thank the anonymous referee for the useful comments. 

%

\vspace{5mm}

\end{document}